\documentclass[pra,amssymb,amsmath]{revtex4}
\usepackage{graphicx}
\usepackage{color}
\newtheorem{thm}{Claim}

\newtheorem{prop}{Proposition}
\usepackage{ulem}

\newcommand{\halv}{\frac{1}{2}}
\newcommand{\bc}{\begin{center}}
\newcommand{\ec}{\end{center}}

\begin{document}

\title{Libertarian free will and quantum indeterminism} \author{Chetan
  S.  Mandayam Nayakar\footnote{aka M.  N. Chetan Srinivas}, S. Omkar,
R.      Srikanth}    \email{srik@poornaprajna.org}
  \affiliation{Poornaprajna   Institute    of   Scientific   Research,
    Sadashivnagar,   Bangalore,  India}   \affiliation{Raman  Research
    Institute, Sadashivnagar, Bangalore, India}

\begin{abstract}
The basic question in the long-standing debate about free will (FW) is
not  whether FW  can  be demonstrated  to  exist nor  even whether  it
exists, but  instead how  to define it  scientifically.  If FW  is not
dismissed   as  an  illusion   nor  identified   with  a   variety  of
unpredictability, then logical paradoxes arise that make FW elusive to
define.  We resolve these paradoxes through a model of FW, in which FW
is  a new causal  primitive empowered  to override  physical causality
under guidance.  We develop  a simple mathematical realization of this
model, that when applied to quantum theory, suggests that the exercise
of FW corresponds to a nonlinear POVM causing deviations from the Born
rule.   In principle, these  deviations would  stand in  conflict with
known conservation  laws and invariance principles,  implying that the
brain,  the  presumed  seat  of   FW,  may  be  an  arena  of  
non-standard physics.   However, in practice  it will be  difficult to
distinguish  these  deviations  from  quantum and  neural  noise,  and
statistical  fluctuations.  We  indicate possible  neurobiological and
neurological  tests,  implications and  applications  of our  proposed
model.
\end{abstract}

\maketitle

\section{Introduction\label{sec:intro}}

Free will (FW),  as we informally understand it in  daily life, is the
power  of a  rational  agent to  pick  her/his own  choice from  among
various  alternative  possbilities.   But  what exactly  is  FW?   The
age-old  question  has provoked  much  debate  among philosophers  and
scientists   \cite{tim,fw,linda,fw3}.   Our   outlook  on   the  world
implicitly assumes that  human behavior is governed by  FW: we choose,
we plan and  we normally hold people responsible for  what they say or
do, either because we imagine that they are at liberty also not to act
so,  or  because we  don't  have  the  freedom to  believe  otherwise!
Neuroscience has been uncovering  causal chains that appear to explain
our  choices, emotions  and even  beliefs, in  terms  of neurophysical
events  extending back  to  the pre-natal  stage.   As much  as it  is
interesting and important to understand whether we have FW, there is a
more elementary  and pressing problem to  deal with: namely  to try to
define FW rigorously and scientifically  in a way that agrees with the
above intuitive  notion.  This last  requirement is important,  as one
can define almost  anything one wants in a theory  of one's own design
\cite{esposito}.  Having defined  FW, one might ask whether  such a FW
is compatible  with known scientific knowledge or  can be demonstrated
experimentally as a new kind of resource in Nature.

Philosophers over  the millenia  have proposed different  responses to
the  problem.   Two  prominent positions  are  \textit{compatibilism},
according   to  which   determinism   is  compatible   with  FW,   and
\textit{incompatibilism},   according    to   which   the    two   are
incompatible.   There   are    shades   of   intermediate   positions.
Compatibilism  is espoused  by, among  others, Calvinists  who believe
that  personal  freedom  to  choose does  not  preclude  foreknowledge
(possessed  by a  all-powerful intelligence)  of future  choices.  Two
broad    incompatibilist    positions    are    \textit{(metaphysical)
  libertarianism}  and \textit{hard  determinism}.   According to  the
former,  in a  situation of  alternative possibilities  (AP)  which is
ontologically (metaphysically) available, an  agent has the liberty to
choose  one  or  other  option,  which is  not  pre-determined.   This
position   upholds  FW,   and  rejects   determinism.    By  contrast,
\textit{hard   determinism}  upholds   determinism  and   rejects  FW,
relegating  it  to  an  illusory feeling.   In  Western  philosophical
thought, this is represented by, among others, Armenianism.

Recently,  a number  of physicists  have  studied FW  mainly (but  not
exclusively)    in   connection    with    quantum   mechanics    (QM)
\cite{ck,ck1,stapp,hoo07,nic,suControl,sugen,suflash,vedral,gisin,
  gisin3440,hall, sabine,svet}.  The important contribution of quantum
mechanics  to this  debate is  in introducing  a concrete  instance of
fundamental indeterminism via the  $|\psi|^2$ Born rule.  Whether this
indeterminism is epistemic or ontic  still remains a moot issue and is
related   to   the  notion   of   realism   in   the  sense   of   the
Bell-Kochen-Specker  theorems.  In  the quantum  approach to  FW, some
have assumed that  quantum indeterminacy gives `elbow room'  for FW to
act, while  others identify FW with  unpredictability, or independence
from  all past information  (with `past'  defined in  an appropriately
relativistically invariant  way).  We call  this approach \textit{soft
  incompatibilism},  according  to   which  FW  is  incompatible  with
determinism but compatible with indeterminism. However, this approach,
which  also   finds  favor  in  the   neuroscientific  and  artificial
intelligence (AI) communities, fails  to capture the sense of control,
of  having liberty  and making  a deliberate  choice, implicit  in the
notion of FW.

In this work, we aim  to define libertarian FW, which encompasses such
a  liberty,  and  to  study  its  relation to  the  laws  of  physics.
Libertarian FW is arguably the  most intuitive version of FW, and yet,
as we will find, the most elusive  to define.  We do not claim that it
is somehow  a better variety of  FW, or that it  can be experimentally
demonstrated at  this time.  It  is our preferred understanding  of FW
simply because  it feels right!   The article is arranged  as follows.
Section \ref{sec:mackay} discusses a specific compatibilist version of
FW, due to the cognitive scientist Mackay.   
Section \ref{sec:wfwp}  presents  a logical  paradox that  would
afflict attempts to define  libertarian FW, and presents a resolution.
Section  \ref{sec:sfwp} explains  another basic  logical  paradox 
about libertarian  FW.  This  logical
barrier is surmounted in Section  \ref{sec:fun} through a new model of
FW.  Certain biological and philosophical aspects of  the model  are  
considered in
Section     \ref{sec:funhash}.      Sections    \ref{sec:fun+}     and
\ref{sec:fun++}  may   be  omitted   by  readers  not   interested  in
quantitative aspects  of the model.  Section  \ref{sec:fun+} presents a
simple  mathematical  representation  of  the  proposed  model,  while
Section \ref{sec:fun++} applies it  to quantum mechanics and considers
some  unusual  consequences.   Neurological  tests,  implications  and
medical applications of  our model are considered  in Section 
\ref{sec:neuro}.  We finally conclude with Section \ref{sec:end},
with a discussion on FW in connection with Darwinian evolution and
the subject of consciousness.

\section{The Mackayan argument: Free Will from uncomputability 
  \label{sec:mackay}}

We mention  two flavors of what are  arguably compatibilist positions:
Calvinist-like  (mentioned  above)  and  Mackayan.   In  the  Mackayan
position \cite{mackay},  the idea here  is that even  in deterministic
dynamics,   self-predictions   lead   to   indeterminacy,   which   is
interpretable as FW.

According to this  argument, given prediction $P$ about  agent $X$, if
$X$  believes $P$,  then  she can  falsify  it by  acting contrary  to
$P$. One might suppose that a more detailed algorithm to arrive at $P$
would be able  to take into consideration $X$'s  reaction, but $X$ can
simply choose to  falsify \textit{that}. It then follows  that only if
$X$  disbelieves $P$,  then $P$  holds with  certainty, but  then this
would make $X$ a kind of inconsistent agent, for disbelieving a truth.
Thus, if  $X$ (and  hence the universe)  is consistent, then  $X$ must
believe $P$. To avoid the first inconsistency, that she falsify $P$ if
she knew it, we conclude that believing $P$ affects $X$'s brain state,
invalidating  the premise that  went into  reaching $P$.   Thus, $X$'s
falsification of $P$ does not  imply inconsistency.  But it does imply
that any  consistent predictive algorithm  $P$ is unable  to encompass
her reaction to  $P$.  Therefore, according to Mackay,  it is upto $X$
how  $X$ chooses to  act; she  can make  $P$ or  $\neg P$  true.  This
indeterminability is interpreted as FW.

A caeful scrutiny reveals the similarity of the structure of the above
argument  analogous  to  G\"odel's celebrated  incompleteness  theorem
\cite{god}, when  the notion of \textit{provability}  is replaced with
that  of   \textit{belief},  though,  strangely,   Mackay  denied  the
connection.   To recast  the Mackayan  position  in a  form closer  to
G\"odel's  theorem: in a  consistent sufficiently  rich, deterministic
world,  there will  be unpredictable  situations. Not  surprisingly on
account of the presence  of self-reference, Mackay's argument can also
be  compared readily  to Turing's  proof of  the unsolvability  of the
halting  problem, where  the putative  algorithm that  outputs  $P$ is
analogous  to the would-be  halting algorithm  $h$ and  the nay-saying
agent  to the  contradictive Turing  machine that  uses the  $h$  as a
sub-routine.  Mackayan  free will, like  Turing uncomputability, comes
from the fact that the cardinality of the set of outcome situations is
greater than proofs of prediction  of an agent's choice.  Thus, in the
Mackayan  argument,  unpredictability  of  a free  agent's  choice  is
identified with G\"odel incompleteness and Turing uncomputability.

But  does incompleteness or  uncomputability constiute  a kind  of FW?
Intriguing  though the  idea  is,  we think  that  it is,  ultimately,
philosophically unsatisfactory,  because it identifies  free will with
unpredictability   rather   than  with   genuine   freedom,  or   even
indeterminacy.  Likewise,  from a philosophical  perspective, we think
that the soft incompatibilist approach from quantum mechanics does not
go far enough, because sense of  control in FW is not explained. FW in
its various quantum avatars looks just like free whim!

\section{The Weak Free Will paradox and its resolution\label{sec:wfwp}}

The  reason libertarian FW  resists definition  is that  attempting to
define it leads to logical paradoxes, which we discuss in this and the
next section.   Although these paradoxes  were not, to  our knowledge,
explicitly  mentioned before,  yet  they implicitly  crop  up in  past
attempts  to  define FW,  thwarting  a  definitive  resolution to  the
problem for over two millenia.  It  is not surprising that in the face
of this daunting logical barrier,  scientists tend to adopt a position
other than libertarianism.

If  we  accept  the   libertarian  position,  then  by  assumption  of
incompatibilism, one rules out  determinism as the universally correct
description of the physical  laws.  But neither does indeterminism (as
governed by some fixed probability rule, $P$) leave enough room for FW
to act.  Consider the sample mean $X_n$ over $N$ trials
$$\lim_{n                      \rightarrow                     \infty}
\textrm{Pr}\left(\left|\overline{X}_n-\mu\right|>\epsilon\right)=0,
$$ by  the Weak  Law of Large  Numbers. This  implies that there  is a
`probability   pressure'  \textit{not}  to   choose  \textit{atypical}
sequences, which  would cause deviations from the  sample mean.  Thus,
there is a kind of long-run determinism, and hence a restriction on FW
as we intuitively understand it.  For  example, given a coin with outcome
space  $\Omega = \{H,  T\}$ and the corresponding
probability vector $P =  (\frac{1}{2},\frac{1}{2})$, the
coin  is not  free to  indefinitely choose  outputs HHHHHHHHH$\cdots$.
Thus  liberatarian  FW  is  compatible with  neither  determinism  nor
indeterminism,  and  thus  belongs   to  the  position  of  {\it  hard
  incompatibilism}.   This  situation  creates   a  paradox   for  the
libertarian  position,  which we  call  the  Weak  FW Paradox  (WFWP),
whereby FW is compataible with neither of the available alternatives.
The other mentioned positions on FW are unaffected by WFWP.

Therefore, if  we accept libertarian  FW, the free-willed  choice will
potentially intefere with the underlying physical dynamics ${\cal D}$.
This interference will  take the form of (a)  overriding causality, if
${\cal  D}$  is deterministic;  or  (b)  causing  deviations from  the
relevant  probability  rule $P$,  if  ${\cal  D}$ is  indeterministic.

Clearly, it is immaterial  whether the underlying physics is classical
or quantum.   The arugment sometimes made,  that quantum indeterminism
gives  `room'  for  free-willed  action   is  thus  not  found  to  be
persuasive.
FW itself can't  be part of the dynamics ${\cal D}$,  for in that case
it could not produce the required deviations.  Therefore, we require a
Cartesian   dualism  with   a  physical   and  \textit{extra-physical}
component making  up a free-willed
 agent.   Physical dynamics ${\cal  D}$ governs
the former while FW comes from the latter.

\section{The Strong Free Will Paradox \label{sec:sfwp}}

The above  resolution of WFWP says  that if libertarian  FW exists, it
must   be  an  extra-physical   resource  whose   intervention  causes
deviations from  ${\cal D}$.  It  still leaves open the  question what
libertarian FW is or how it  can be consistently defined. Here we will
show how trying to pin down its nature leads to another paradox, which
we call the Strong Free  Will Paradox (SWFP).  The paradox is obtained
essentially by extending the traditional incompatibilist claim for the
incompatibility    of    determinism    and    FW,    to    that    of
\textit{indeterminism}  and  FW.   The  idea  is  that  the  truth  of
determinism or indeterminism would  mean that we don't control actions
in  a way one  would expect  of self-determining,  free-willed agents.
Since  determinism and  indeterminism  are the  only available  causal
primitives, there  is no  such thing as  libertarian FW,  according to
this line of reasoning.  There is  a similarity to WFWP, but the focus
has shifted from ``where?'' to ``what?'', or even ``whether?''.

We present two  slightly different versions of SFWP.   In the first of
them, one  assumes that FW exists and  tries to locate it  in terms of
properties of the agent.  According to this argument:
\begin{description}
\item{\bf  [I]} An  agent  has  free will  only  when her  intentional
  actions emerge from the agent herself.
\item{\bf  [II]} Therefore,  they are  deterministic functions  of her
  volitions,    beliefs,   desires   etc,    which   we    denote   by
  \textit{variables of intent}, $x^{(1)}_j$.
\item{\bf  [III]} If  there  are  no prior  causes  of her  volitions,
  desires, etc., then  the values $x^{(1)}_j$ take on  must be random,
  making the agent's volitions,  desires, etc., and hence her actions,
  whimsical.
\item{\bf [IV]} Since whimsicality undermines the notion of control or
  intent, $x^{(1)}_j$ must not  be random, but deterministic functions
  of some  other second order intent variables  $x^{(2)}_k$, which, by
  virtue of \textbf{[I]}, must belong to the agent herself.
\item{\bf  [V]} As  before, the  variables $x^{(2)}_k$  themselves are
  caused  or uncaused.  If  the latter,  then $x^{(2)}_k$  are random,
  making   $x^{(1)}_j$,  and  hence   also  her   actions,  ultimately
  whimsical.  But if  the former,  then $x^{(2)}_k$  are not  free but
  depend,  by   a  similar   reasoning,  on  higher   order  variables
  $x^{(3)}_j$, and so on similarly to even higher orders indefinitely.
\item{\bf [VI]} If this pattern of recursion terminates at some finite
  depth $N$, then $x^{(N)}_j$ either  has no causes or external causes
  of  unspecified origin. In  the former  case, we  obtain capricious,
  indeterministic behavior, in  the latter case, unfree, deterministic
  behavior. Neither connotes FW as we recognize it.
\end{description}
In brief, when we try to incorporate the notion of will or intent into
the  action,  the  action  becomes  deterministic  and  hence  unfree.
Putting  freedom  back means  removing  determinism, which  undermines
intent by making  the action random and the  agent whimsical. Thus the
`free' and  the `will' in `free  will' are at  loggerheads, making the
word  an  oxymoron. This  is  the  Strong  Free Will  paradox  (SFWP),
according  to which  randomness and  determinism seem  to be  the only
fundamental causal primitives in Nature, with libertarian FW a figment
of imagination.

It is of  interest to note that certain classical  accounts of FW fall
prey  to SWFP,  illustrating  its  elusiveness to define.
According to  Thomas Hobbes, ``A  free agent is  he that can do  as he
will,  and forbear  as he  will....''.   David Hume  
characterizes it thus: ``power  of
acting or of  not acting, according to the  determination of the will:
that is, if we choose to remain at rest, we may; if we choose to move,
we also  may....  This hypothetical liberty is  universally allowed to
belong  to  everyone  who  is  not  a prisoner  and  in  chains."   To
paraphrase in terms of our discussion above, they both are essentially
saying  $E =  E(x_j^{(1)})$, but  weren't taking  the  argument farther.
Arthur Schopenhauer does take it  one step farther but says: ``You can
do what  you will,  but you cannot  will what  you will. In  any given
moment  of  your  life  you  can  will only  one  definite  thing  and
absolutely nothing other than that one thing.'' He is thus effectively
responding to  SWFP by characterizing 
$E  = E\left(x_j^{(1)}(x_k^{(2)})\right)$ and
denying the existence  of libertarian FW. One can  try other variants,
such  as $E =  E(x^{(1)}_j)$ being  a probabilistic  function, whereas
$x^{(1)}_j = x^{(1)}_j(x^{(2)}_j)$  being deterministic, so that there
is  mix   of  determinism   (Will)  and  indeterminism   (freedom)  at
\textit{different}  levels of  the agent's  personality.  However, the
core of SFWP remains.

A related, but different version of SFWP, tries to identify FW as a
particular kind of influence over the act of making a choice. 
According to this argument:
\begin{description}
\item{\bf [A]}  Suppose that from  a set of  alternative possibilities
  (AP), a choice is eventually made by an agent.
\item{\bf [B]} From  the principle of excluded middle,  the choice was
  made either  according to a rule or  it was not made  according to a
  rule.
\item{\bf [C]} In  the former case, there is  no genuine AP situation.
  Hence  we  have  determinism,  and  there  is  no  FW.
\item{\bf [D]} In the latter case, we have pure randomness, and hence,
  again,  no FW  in the  sense of  the agent's  control  and voluntary
  choice. 
\end{description}
We thus find again that the agent's choice is deterministic or random,
with no apparent room for (libertarian)  FW.  This is the basis of the
\textit{pessisimist}  view that  there  can  be no  such  thing as  FW
\cite{fw}.

Implication  {\bf   [C]}  lies  historically  at  the   heart  of  the
incompatibilist argument.  Here, if the rule is \textit{prescriptive},
we have  {\it causal  determinism}, as a  law of physical  dynamics in
Nature.   If  the rule  is  only  \textit{descriptive},  we have  {\it
  logical  determinism},
with outcomes  of  future choices being  known to  an
all-powerful intelligence and assigned definite truth values.

\section{Resolution of the Strong FW paradox \label{sec:fun}}

We propose  the following hard incompatibilist  model as a  way out of
SFWP, drawn  from an interpretation of  Eastern philosophy.  According
to  the model,  the conscious  personality of  the  free-willed agent,
which  may be  called the  \textit{Ego}  (after the  father of  modern
psychoanalysis, Sigmund  Freud), is influenced  in an AP  situation by
three structural  elements.  This tripartite division is  at the heart
of  our model.   Two  of the  elements,  and their  functions, are  as
follows:
\begin{description}
\item{\bf  Nature (N).}   Imposes mental  constraints in  the  form of
  desires, instinctive drives and emotional tendencies.

\item{\bf Understanding (U).}  Offers guidance in action through
a rational capacity to model the world,
  and understand the  (ethical, social, financial, etc.)  implications
  of each choice.  
\end{description}

When a situation with alternative possibilities is encountered, Nature
{\bf   N}  and   Understanding  {\bf   U}  present   their  respective
recommendations  to the  Ego  on selecting  one  of the  alternatives.
While the recommendation of  Nature appears as a \textit{desire}, that
of  Understanding appears as  a \textit{thought}  or \textit{feeling}.
As  a specific  example, the  desire may  be oriented  towards seeking
pleasure (Freud's  pleasure principle),  while the thought  or feeling
may   be  about  conscientious   restraint.   The   dilemma  sometimes
experienced by  one about  to make a  choice is the  possible conflict
between the two recommendations.

It is convenient to designate  the \textit{mind} as the seat of Nature
{\bf  N}.  In  Freudian psychoanalytic  terminology, the  mind  may be
identified with the \textit{id}.  Similarly,  the seat of ${\bf U}$ is
posited  to be the  \textit{intellect}, which  may be  identified with
Freudian  \textit{superego}.   To use  a  crude  but visually  helpful
computer  analogy,  we  may  picture  the  ego  as  the  CPU  (central
processing  unit)  of  a compter,  with  {\bf  U}  and {\bf  N}  being
softwares  uploaded  onto  it  during  the waking  period  from  their
respective seats, which are like hard disk memory locations.

For  non-libertarian accounts of  FW, including  AI, the  two elements
{\bf U} and {\bf N}  suffice. They are both \textit{causal resources},
i.e., systems  that can drive the agent's  behavior independently, and
thus aspects of the dynamics  ${\cal D}$.  The eventual choice will be
some  deterministic  or  probabilistic  outcome  of  their  interplay.
Philosophically speaking, the main drawback with such accounts is that
they lack the sense of having liberty to make a deliberate choice.

Crucially, for (libertarian) FW, {\bf U} is not a causal resource, but
a  \textit{guidance resource}, unlike  {\bf N}.   When the  input from
{\bf U} and {\bf N} are experienced as possibly conflicting tendencies
by  the Ego,  the guidance  by itself  is powerless  to  influence the
agent's choice.  A further element  is required, the third and last in
our  model,  which  expresses  the  idea  of  empowering  the  Ego  to
deliberately  choose   between  the  recommendations   of  desire  and
guidance.  This is the faculty of volition, or simply:
\begin{description}
\item{\bf Freedom  of Will (F).}   The \textit{extra-physical} freedom
  to orient or  align the choice in line  with the Understanding ${\bf
    U}$,  by  overcoming, if  necessary,  the  Constraints imposed  by
  Nature ${\bf N}$.
\end{description}
The extra-physicality  follows from the  resolution of the  WFWP. SWFP
also implies  it because if it  were not extra-physical,  then {\bf F}
would be part of the underlying dynamics ${\cal D}$ as a deterministic
or indeterministic aspect, and not libertarian, as required. As to the
question of  whether such an extra-physical resource  is sanctioned by
physics or can be tested,  we return to it in Section \ref{sec:fun++}.
Here   we    are   concerned   only   with    obtaining   a   suitable
\textit{definition}.  We call our  model `FUN'  in recognition  of the
fact that it involves the triad  of elements {\bf F}, {\bf U} and {\bf
  N}.

Faced  with an  AP  situation, Nature  and  Understanding may  present
conflicting  recommendations to  the  Ego under  some description--  a
desire  or   drive  vs   a  thought,  feeling   or  belief.    If  the
Nature-induced desire is  aligned away from the Understanding-proposed
thought,  then supported  by  FW,  the thought  acts  as a  restraint,
whereas if the desire is  attuned to {\bf U}'s recommendation, then FW
acts  to that  extent  effortlessly.   Together {\bf  F}  and {\bf  U}
constitute  a  causal  resource,  in  which {\bf  F}  contributes  the
magnitude or ``energy'' while  {\bf U} the direction. Recognizing this
is crucial in  resolving SFWP, and its oversight is  the cause of much
confusion in  existing accounts  of FW. Thus  FW is the  expression of
{\bf F}  in a situation  where the ego  is exposed to  the potentially
conflicting recommendations of {\bf U} and {\bf N}.

FW  as   defined  above  is  the  new   \textit{causal  primitive}  or
\textit{principle   of   causation},   apart  from   determinism   and
indeterminism.  By  its  fundamental  character, it  is  empowered  to
override  physical   causality  (of  Nature)   under  the  intellect's
guidance.   FW  thus  enables  a  state  of  affairs  wherein  logical
determinism holds whereas causal determinism fails.  We return to this
point elsewhere to use this  as a basis for a (meta-)logical framework
to  characterize  FW.  We  think  that our  definition  of  FW, and  a
quantification   of    it   below,   agree   with    the   notion   of
\textit{volitional  causation} proposed  in  Ref.  \cite{hodgson},  as
distinguished from \textit{physical causation}

The agent's freedom to control, together with the Understanding, as we
have   proposed,  bears  some   similarity  to   the  \textit{Ultimate
  Responsibility}  posited   by  the  philosopher   Kane  \cite{kane}.
However, the role of {\bf F} and {\bf U} need not be confined to moral
issues alone,  and may  extend to other  walks of life.   For example:
agent  $X$  is   suffering  from  a  medical  problem,   and  has  the
Understanding  that $X$  needs  to undergo  treatment.  However  $X$'s
Natural tendency, governed  by Constraints {\bf N}, would  be to avoid
the treatment because it would  cause discomfort.  $X$ exercises FW to
endure  the discomfort  in the  interest  of later  cure and  longterm
benefits.

We will now prove that the FUN model resolves SFWP. In particular, the
above example highlights  how the \textit{incompatibilist} implication
{\bf [C]} of SFWP fails.  If  one happens to be familiar with the {\it
  Weltanschauung},  and hence the  Understanding of  $X$, and  $X$ has
high free  will, then the  \textit{freely} chosen alternative  will be
\textit{predictable}.  Proposition  \ref{pro:saint} below, whose truth
is clear in  this light, is rigorously validated  by our definition of
FW.  We  define a  \textit{Saint} as an  agent whose  Understanding is
ethical and whose FW is near maximal.
\begin{prop}
Presented with  the choice  between the good  and the evil,  the Saint
{\rm freely} chooses the good.
\label{pro:saint}
\end{prop}
By definition, the  Saint, equipped with maximal FW,  can if necessary
override any Constraints like desire and instinctive drives imposed by
his human nature, and align his choice in tune with his Understanding.

Thus, the deterministic behavior  and predictability of the Saint does
not preclude  his free will, contrary  to the \textit{incompatibilist}
implication  {\bf  [C]}  in SWFP. Knowing that a person is a Saint,
we have sure \textit{foreknowledge} of the ethicality of all his future 
choices, \textit{because} he is endowed with high FW.
 We  have  here  an instance  of  a
descriptive    rule   (logical    determinism)   that    is   strictly
non-prescriptive    (i.e.,   strictly    not    causal   determinism).
Intuitively, this is clear, since one should be able to choose freely,
even to act predictably.  Not to  be able to choose to act predictably
would imply a constraint on freedom.

The  instance   of  Saintly  determinism  does  not   make  our  model
compatibilistic, as it must  be clear. In particular, other situations
are  allowed  to  exist  by  our model,  that  resist  compatibilistic
interpretation,  as noted  below. Suppose  that an  agent's FW  is not
maximal.   Then his  choice will  fluctuate randomly  between choosing
according to the dictates of  his nature and the recommendation of his
Understanding.   This  is illustrated  in the  following
proposition, whose  intuitvely apparent truth  is validated rigorously
by our definition of FW.
\begin{prop}
Presented with the  choice between  the good  and the  evil, 
the Conscientious Criminal vacillates.
\label{pro:cc}
\end{prop}
This criminal,  being conscientious, has a clear  Understanding of the
virtue of  ethical behavior, but,  owing to lack of  sufficiently high
FW, cannot always overcome the  compulsion of his criminal nature. His
choice is random, being good  sometimes and being evil at other times.
Thus the  randomness of his choice  does not imply lack  of free will.
Here it only  implies low free will.
The \textit{pessimist} \cite{fw} implication {\bf [D]} in SWFP is also
therefore falsified. The recognition of FW as a new form of causation
also clarifies why the first version of SWFP fails. For example,
implication \textbf{[VI]} is incorrect for reasons stated above.

In contrast  to Proposition \ref{pro:saint},  predictable behavior can
arise also because of low degree of FW.  Consider:
\begin{prop}
Presented with the  choice between  the good  and the  evil, 
the Hardcore Criminal chooses evil.
\label{pro:hcc}
\end{prop}
Although the Hard-core Criminal  may possess an ethical Understanding,
yet,   because  he   almost   entirely  lacks   FW,   his  choice   is
deterministically  decided by  his evil  nature. Knowing that a
person is a Hardcore Criminal, we have foreknowledge of the iniquity
of all his future choices because he \textit{lacks} any FW.  
If  we theoretically
empower  him with  some  FW,  we obtain  the  scenario of  Proposition
\ref{pro:cc}.   In  other  words,   adding  a  degree  of  FW  removes
determinacy.  Clearly, this  prediction of the model is  not in accord
with  compatibilism. Therefore, our  model is  neither incompatibilist
nor compatibilist.

\section{Some neuroscientific considerations and
philosophical musings \label{sec:funhash}}

The  FUN model  as it  stands  does not  explain the  notion of  moral
responsibility.   When the  given  an  AP situation  is  a moral  one,
responsibility arises arguably because a  person has the power {\bf F}
to follow through  the guidance provided by {\bf  U} by overcoming the
dictates  imposed by  {\bf  N}. The  model  does not  explain how  the
differences in the degree of freedom, eg., between that of a Saint and
a  Conscientious Criminal,  arise.   Unless an  agent  can somehow  be
construed as having played a role in the current level of freedom, the
judgment of moral  responsibility must be relative (to  a given degree
of freedom). Further assumptions are needed, which will be taken up in
a  subsequent  work,  to  develop  a more  complete  notion  of  moral
responsibility in the model.

For  the triad  of  elements of  the  FUN model  described in  section
\ref{sec:fun},  we suggest the  following plausible  brain correlates.
For convenience,  we refer to  this neurologically adapted  version of
FUN as  the FUN\# model.   The element {\bf  N}, and hence  the mind's
brain  correlate, is  probably  the limbic  system, which,  comprising
brain  structures like the  hippocampus, amygdala,  etc., is  known to
support a  variety of functions, including behavior,  emotion and long
term memory.   The element {\bf  U}, and by extension  the intellect's
brain correlate is located  presumably in the brain prefrontal cortex.
This  region   of  the  brain  is  believed   by  neuroscientists  and
psychologists to  be responsible for many  higher cognitive functions,
such  as   planning,  differentiating   between  the  good   and  bad,
determining consequences of actions, planning actions, and so on.

The FUN\#  model posits that,  while these two regions  participate in
decision   making   processes,  they   do   not   correspond  to   the
\textit{ultimate  control module}  (UCM), the  brain correlate  of FW.
This could be a localized region, such as the pineal gland.  The claim
for this gland is mainly historical. The FUN\# model accepts this as a
working hypothesis on the strength  of the observation that this gland
is centrally located in the  brain, as befits a UCM.  Physiologically,
this  would mean that  brain areas  corresponding to  voluntary muscle
movement (the motor cortex) will be functionally linked to the frontal
cortex, the limbic  system and the region of  the pineal gland, during
the execution of planned actions.

In  the FUN  model,  {\bf F},  {\bf  U} and  {\bf  N} are  independent
elements that interact when a  choice is made.  
In actual fact, they influence
each    other   at    the    \textit{subconscious} level  in    complicated
situation-dependent ways, so  that the elements as they  appear at the
conscious level are `dressed versions' rather than the `bare versions'
(to use  terminology borrowed from  optics or particle  physics).  The
agent's personality  is thus a  nonlinear functional of  the elements,
where each  element influences the  others and also  itself indirectly
through its effect on the others.

An  instance  where {\bf  U}  affects {\bf  N}  is  in consumers'  brand
loyalty,   which  manifests  as   preferences,  detectable   as  neural
correlates, that  humans display  towards products that  are otherwise
similar (cf.  an interesting study reported in Ref.  \cite{brand} with
regard to drinks).  An instance where {\bf N} affects {\bf U} is 
the familiar case of
people  showing reduced objectivity  when dealing with  those
they have  a strong  emotional connection  (whether  negative or
positive) with.  An instance where  {\bf N} affects (diminishes) freedom is
when a  drug addict (resp.,  smoker) relentlessly seeks his  next shot
(resp., puff) even though he knows it is not good for him; and so on.

For  this reason,  from a  behavioral  perspective, it  would be  more
convenient to  talk not in terms of interaction between the elements,
but instead in terms of interaction between three broad traits:  
that dominated  by Nature {\bf  N} (``the  animalistic"),  
that  dominated  by  freedom  (``the saintly"), and   
that wherein these two are in balance  (``the  human").
The  philosophy  of \textit{Yoga}  \cite{srippisr}  terms
these   three   basic   traits  as \textit{tamas},   \textit{rajas}   and
\textit{sattva}.   In  this line  of  thought,  a  human character  is
constituted  by  these three  basic  traits  and  determined by  their
relative dominance.   The \textit{sattvic}  type  of humans
have  the  highest  degree   of  FW,  while  the  \textit{tamasic}
have the lowest.

\begin{figure}
\includegraphics[width=15cm]{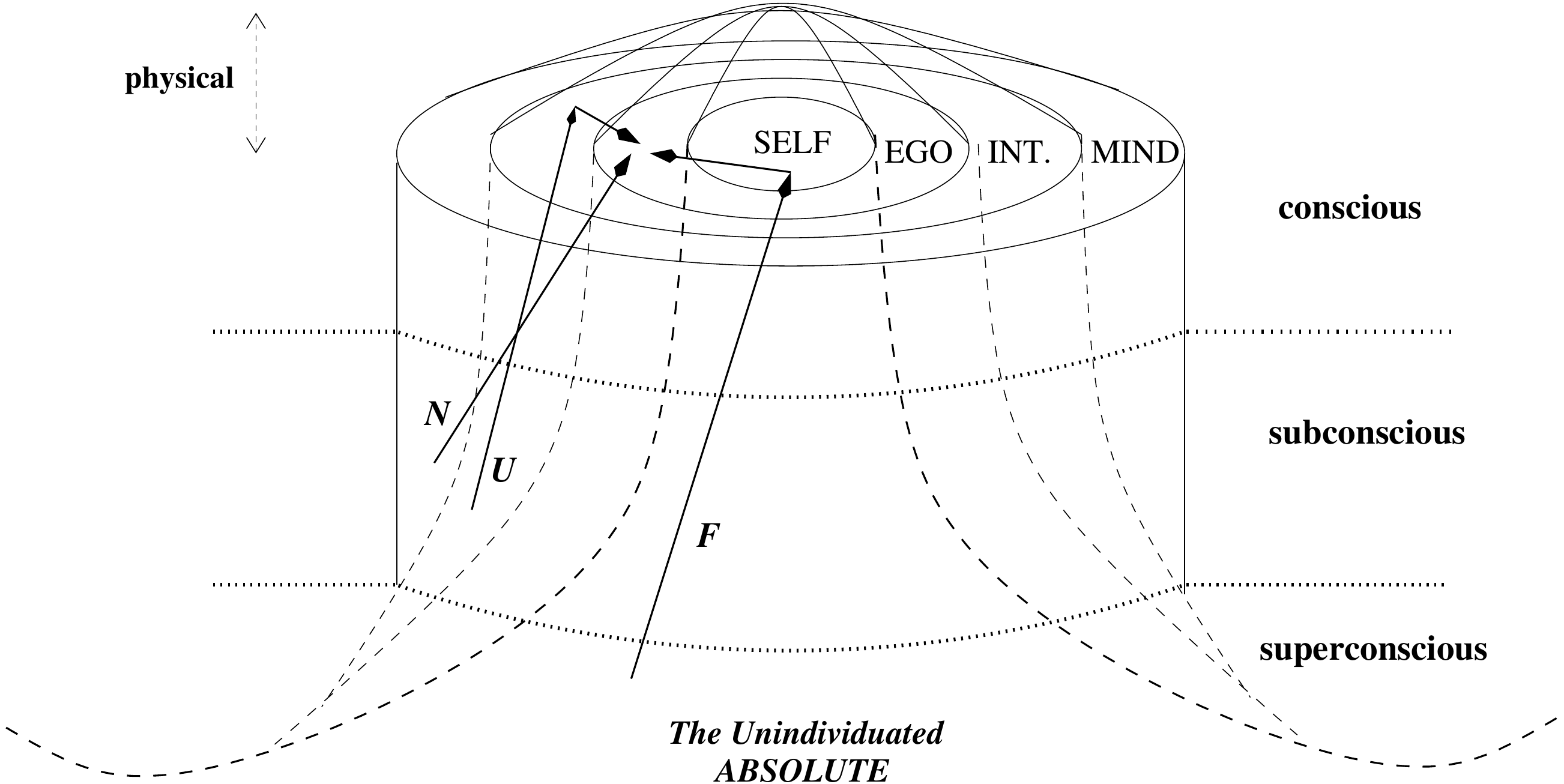}
\caption{Scheme of the  structure of the individual psyche  as used in
  our model of  FW.  The scheme is based primarily  on the the Eastern
  philosophy  of  \textit{Vedanta}  \cite{srippisr} and  \textit{Yoga}
  \cite{srippisr},   with  inputs  from   Freud's  iceberg   model  in
  psychoanalysis and  E. Cayce  \cite{ec}. The ego  receives causal
  constraints from  the mind, guidance from the  intellect and freedom
  from  the  Self  to  direct  the  choice  in line with the  guidance.
  In a word: \textit{Thought over Nature}.   In  this  geometric
  interpretation,   the  ego   consciousness   is  \textit{horizontal}
  (spreading out  concentrically from the  center), while the  mode of
  Cognizance  of the Self  is the  \textit{witness consciousness}  
  (\textit{s\'aksh\'i bh\'ava} in Sanskrit). This represents a bottom-up
  perspective  from  the  base of the figure,  which  is
  \textit{vertical} and  illumines all overlying elements of the
  personality, including  the ego. In this sense, the Self's
  Witness consciousness  is higher dimensional  than the personality's
  ego consciousness.  (An even higher dimensional consciousness can be
  associated geometrically with an observer looking at the above image
  from outside the page. 
  This perspective represents a subtler, Witness of Witness
  Consciousness.  And  so on.  Vedantic philolosophy  posits that this
  hierarchy   terminates   in   \textit{brahman},  the   Absolute   or
  the fundamental subtratum.)}
\label{fig:isbergr}
\end{figure}

To complete  the model,  we offer some  thoughts on  the extraphysical
agency  to which  the will  is ascribed.  Here we  will appeal  to the
philosophies   of   \textit{Vedanta}   \cite{srippisr} and Yoga,
which recognize  and  supply such  an
entity--  the \textit{Self} (\textit{\'atman}  in Sanskrit),  which is
quite distinct from the  Ego (\textit{ahamk\'ara}) and conceived of as
the essential individual. At a deeper level of consciousness, the Self
fans out to serve as  the overarching \textit{substratum} on which the
agent's  faculties  of Ego,  Understanding  and  Nature  rest. At  the
deepest  level  of  consciousness,  there  is  no  structure  and  the
individual      Selves     resolve     into      an     unindividuated
Absolute.  Contradictory  as  these  descriptions seem,  they  can  be
readily        interpreted       geometrically,        cf.       Figure
\ref{fig:isbergr}.   As in  the   Freudian  \textit{structural
  iceberg model}, the Constraints of {\bf N} are communicated from the
subconscious mind to the conscious Ego, where they are experienced
as desires. The Understanding  and Will are exercised 
\textit{preconsciously}, in that although
having deeper roots, they are supplied from within the conscious level.

The Self cognizes  the faculties of Ego, mind  and intellect through a
mode   of    Consciousness   called   \textit{witness   consciousness}
(\textit{s\'akshi bh\'ava}).  This is  subtler than and different from
the  agent's Ego  consciousness.  Figure  \ref{fig:isbergr}  depicts a
simple geometric interpretation of these modes of awareness.  
The basic information about
the the above three elements are summarized in Table \ref{tab:3}.

\begin{table}
\begin{tabular}{|c|c|c|c|c|}
\hline ~~~~~{\bf Faculty} & ~~~\textbf{Seat} & ~~~ \textbf{Function} &
~~~~ \textbf{Possible physical correlate } &
\textbf{Character trait} \\ \hline Nature, {\bf N} &
Mind/Id & Mental constraints & limbic system & \textit{tamas} 
\\ \hline Understanding {\bf
  U} & Intellect/Superego  & Guidance & Brain frontal  cortex &
  \textit{rajas} \\ \hline Freedom
{\bf F} & \textit{Self}/~~-- & Control & pineal gland? & 
\textit{sattva} \\ \hline
\end{tabular}
\caption{The three structural elements  of libertarian FW according to
  the FUN model, their corresponding  place of origin (with a Freudian
near-equivalent term), the function of
  the  elements, possible  physical correlates  in the  brain,
  and (in the terminology of Yogic philosophy)  the
  corresponding character trait.}
\label{tab:3}
\end{table}

\section{A quantitative version of the \textit{FUN} model \label{sec:fun+}}

We develop a minimal mathematical model, denoted FUN+, to quantify the
qualitative ideas developed above as  part of the FUN model.  Given an
AP  situation, and  a  choosing event  ${\bf  e}$, let  the choice  be
represented by  a random  variable $E$ over  the sample  space $\Omega
\equiv  \{e_1,  e_2,\cdots,  e_n\}$,  representing  the  possible  $n$
choices that can be made.   If ${\bf e}$ involves a material particle,
which presumably lacks any libertarian  FW, the outcome $e_j$ at event
{\bf e} would occur with probability $p_j$, and there is no distortion
of $P$.

The  dictates  imposed  by  the  Constraints of  Nature  {\bf  N}  are
represented,   over  $\Omega$,   by  the   probability  vector   $P  =
\{p_1,\cdots,p_n\}$, normalized  so that $\sum_{j=1}^n  p_j=1$.  The
vector could be pure  ($\forall_j \left(p_j\right)^2 = p_j$) or mixed.
The recommendation  due to Understanding {\bf U}  is represented, over
$\Omega$,  by the probability  vector $P^U  = \{p^U_1,\cdots,p^U_n\}$,
which is normalized and could be pure or mixed.  
For example, if $\Omega  = \{{\tt
  good}, {\tt evil}\}$,  then we would have $P^U  = \{1,0\}$ according
to the  moral criterion,  whereas by Nature,  $P=\{\halv,\halv\}$.  If
$\Omega   =  \{{\tt   coffee},  {\tt   tea},  {\tt   alcohol}\}$,  the
Understanding  could   use  a  health  criterion  to   return  $P^U  =
\{\frac{1}{2},\frac{1}{2},0\}$, whereas it may be that
$P=\{\frac{1}{4}, \frac{1}{4}, \halv\}$.

FW  {\bf F},  as freedom  to drive the choice from  the  prescriptive $P$
towards the recommended $P^U$,  is represented by the
scalar quantity  $\sigma$ (where  $0  \le \sigma  \le  1$)
such that greater freedom connotes larger $\sigma$.  The  probability
vector $P^\prime$ representing the  eventual choice of an alternative,
is  obtained by distortion of $P$ towards  $P^U$ in the
measure of strength of $\sigma$. A  particularly simple
form to  capture this idea is  a convex combination of  $P$ and $P^U$,
parametrized by $\sigma$:
\begin{equation}
P^\prime  = \sigma P^U + (1-\sigma)P.
\label{eq:ccfw}
\end{equation}
Thus  the random variable  ${\cal P}^\prime$  representing the selected
option is a weighted mean  of random variables ${\cal P}^U$ and ${\cal
  P}$,  representing, respectively,  Understanding  and Nature.   When
$\sigma=1$ (maximal  FW), $P^\prime=P^U$, i.e., the  choice is aligned
with the Understanding.   When $\sigma=0$ (vanishing FW) $P^\prime=P$,
i.e., the choice  is entirely determined by Nature.  
For example, if $P$ is the probability for
obtaining  an outcome  0  or  1 when  measuring  the Pauli  observable
$\sigma_z$  on  a  quantum  two-level  system  (qubit)  in  the  state
$\alpha|0\rangle   +    \sqrt{1-|\alpha|^2}|1\rangle$,   then   $P   =
(|\alpha|^2,  1-|\alpha|^2)$.   Since  the  qubit  arguably  lacks  FW
($\sigma=0$), $P^\prime  = P$ according to  Eq.  (\ref{eq:ccfw}).  The
nonlinear functional  $P^\prime[P]$ must be  viewed as a set  of ontic
probabilities, and thus a fundamental limitation on the predictability
of  an agent's  choice, and  not  epistemic  probabilities
arising from ignorance about the details of an agent.

The  unpredictability $\xi$  of  an  agent may  be  quantified by  the
Shannon entropy of $P^\prime = \{p_x^\prime\}$:
\begin{equation}
\xi \equiv H(P^\prime;\sigma) = -\sum_x p_x^\prime \log p^\prime_x 
= \langle \log(\sigma p^U_x + (1-\sigma)p_x)\rangle,
\label{eq:unpred}
\end{equation}
where  $H(P^\prime;\sigma)$  is  the  Shannon  entropy  of  $P^\prime$
for a  given value of $\sigma$.  The  following two results
quantitatively   present   our earlier  resolution   of   SFWP.
In  each case, we  provide an example in  the following
two theorems, that contradicts each of implications {\bf [C]} and {\bf
  [D]} of SWFP.
\begin{thm}
The predictability  of a Saint's behavior  does not imply  his lack of
FW.
\label{thm:saint}
\end{thm}
{\bf Proof.} We construct an explicit instance of predictable behavior
with  (high)  FW.   Let   $\Omega=\{{\tt  good},  {\tt  evil}\}$.   By
definition, the Saint of Proposition \ref{pro:saint} is ethical in his
Understanding,  so that  $P^U=(1,0)$, and  his will  is free,  so that
$\sigma  \approx  1$.   Even  if  he  may bodily  not  be  attuned  to
perfection, still by dint of his  high FW, he is always able to choose
in accord  with his  (ethical) Understanding, i.e.,  $P^\prime=P^U$ by
Eq.    (\ref{eq:ccfw}).    From   Eq.   (\ref{eq:unpred}),   we   have
$H(P^\prime;\sigma)   \approx  0$   bit,   implying  almost   complete
predictability, irrespective of $P$.  \hfill $\blacksquare$
\bigskip

This implies, as noted earlier, that our model is not incompatibilist.
More generally, we note that $H(P^\prime;\sigma)$ is not a 
monotonous function of $\sigma$. The regime where
\begin{equation}
\frac{dH(P^\prime;\sigma)}{d\sigma} < 0,
\label{eq:sigless}
\end{equation}
 and thus 
increase in FW leads to certainty, may be regarded as a zone
of disagreement with the incompatibilist position. For example,
given a situation involving two choices, with $\Omega =
\{0,1\}$ and $P^U=\{1,0\}$ and $P=\{0,1\}$, 
$dH/d\sigma < 0$ for $0.5 < \sigma \le 1$.

\begin{thm}
The randomness  of the Conscientious Criminal's choice  does not imply
his lack of FW.
\label{thm:cc} 
\end{thm}
{\bf  Proof.}  We construct  an explicit  instance of  random behavior
with non-vanishing  FW.  In the  case of the Conscentious  Criminal of
Proposition \ref{pro:cc},  who has  non-vanishing but not  maximal FW,
let $\sigma=\frac{1}{2}$, and let  $P \approx (0,1)$, corresponding to
the constraint imposed by his evil Nature.  From Eq.  (\ref{eq:ccfw}),
we have  $P^\prime = (0.5,0.5)$,  and hence $\xi \approx  1$, implying
near maximal  randomness in choice. However,  he does not  lack FW, as
$\sigma > 0$.  \hfill $\blacksquare$
\bigskip

In the case of the Hard-core Criminal of Proposition \ref{pro:hcc}, we
have  the same  $\Omega$, $P$  and  $P^U$ as  in the  Example used  in
Theorem  \ref{thm:cc},  but  $\sigma\approx0$, corresponding  to  this
criminal's  low  FW.  From  Eq.   (\ref{eq:ccfw}),  we have  $P^\prime
\approx  P$, and thus  from Eq.   (\ref{eq:unpred}), $\xi  \approx 0$,
implying  very  little unpredictability  in  choice,  similar in  this
respect to  the Saint of Proposition  \ref{pro:saint}. However, unlike
with  the saint,  whose predictability  comes from  high FW,  here the
predictability  is due the  criminal's sure  surrender to  his natural
instinctive constraints. Augmenting his FW, we find that
\begin{equation}
\frac{H(P^\prime;\sigma)}{d\sigma} > 0,
\label{eq:sigmore}
\end{equation}
that is, increase in FW leads to increase in uncertainty
(when $0 < \sigma < 0.5$).
Thus our model is also not compatibilist.
Together, Eqs. (\ref{eq:sigless}) and (\ref{eq:sigmore})
imply that our model is neither compatibilist nor incompatibilist.

The FUN model implies 
that $\sigma$ must be at least partially extra-physical and cannot  be
described by  any purely physical theory,  even a theory-of-everything
(ToE),  that  encompasses  only  physical phenomena.   In  particular,
artificial  intelligence   (AI),  which  is   fundamentally  built  on
(quantum)  physical  rules, cannot  capture  true cognitive  behavior.

These considerations entail  that a human agent, and  by extension any
sentient  agent, could  not  be considered  merely  as a  sufficiently
complex robot,  but a qualitatively distinct class  of entities.  Here
Penrose's  interesting  thesis is  worth  noting,  according to  which
conscious processes are fundamentally non-algorithmic \cite{pen94}.

\section{Quantum indeterminism and free will \label{sec:fun++}}

WFWP  shows that  quantum  indeterminism  is not  better  for FW  than
classical  determinism. However, the  fact remains  that the  world is
fundamentally a quantum mechanical place. Further, there is in a sense
a  lesser departure  from  the physical  dynamics  ${\cal D}$  through
free-willed  intervention if  ${\cal D}$  were  indeterministic rather
than  deterministic,  in that  there  is  only a  \textit{statistical}
violation  of   causality  in  the   former  case,  rather   than  the
\textit{logical} violation of causality, as it is with the latter.

Accordingly,  if  $X$  is  the  random  variable  corresponding  to  a
FW-influenced quantum measurement ${\cal M}$ on system $S$ (presumably
a  suitably small  brain element)  in state  $|\psi\rangle$,  then $X$
would   deviate   from  the   Born   probability  rule.    Free-willed
intervention  will therefore manifest  as {\it  statistical deviations
  from the  Born rule}.   By contrast, the  freedom or `free  will' of
quantum particles, which is  plain quantum randomness, will conform to
the Born rule.  In the  model described below (called FUN++), which is
a  particular  realization of  the  FUN+  model,  we propose  that  FW
intervenes by controling the collapse of the wave function.

To begin with, we represent the
AP  situation as  a  \textit{uniform} quantum  superposition, and  the
willful  choice by  a  directed collapse  of  the wavefunction.   This
quantum model  thus presumes objective collapse  of the wave-function,
and is compatible with interpretations of quantum mechanics that admit
it.   The exercise  of  FW is  broadly  divided into  three stages  as
follows:
\begin{description}
\item[Attention.]  Faced with an AP situation of alternatives $j$, the
  brain creates a quantum  superposition that reflects the dictates of
  Nature {\bf N}:
  \begin{equation} 
|\Psi\rangle = \sum_j  \sqrt{p_j}|j\rangle
  \label{eq:C}
  \end{equation}
 in a suitable  subneuronal system $S$ in an  appropriate basis ${\cal
   B} = \{|j\rangle\}$, where the states $|j\rangle$ correspond to the
 base choice space  $\Omega$.  The main requirement is  that it should
 be  possible to  shield $S$  indefinitely from  decoherence  or other
 noise, while the process of  making a choice is under way.  According
 to Ref.   \cite{penham}, brain microtubules  may be the seat  of such
 superpositions.

\item[Survey.]  The ego surveys  the recommendations  of the  mind and
  intellect.   This   event  is  mathematically   represented  by  the
  preparation  of  a  nonlinear positive  operator-valued  measurement
  (POVM)
\begin{equation}
M_j \equiv  \sqrt{p_j^{-1}\left(\sigma p^U_j  + (1-\sigma)p_j\right)}
|j\rangle\langle j|,
\label{eq:U}
\end{equation}
which satisfy the completeness condition $\sum_j \langle \Psi|M_j^\dag
M_j|\Psi\rangle = 1$.  In the absence of FW,  $\sigma=0$, and the POVM
reduces to ordinary projective measurement.

\item[Collapse.] Application  of the POVM to  the state $|\Psi\rangle$
  causes the transition:
\begin{equation}
|\Psi\rangle       \longrightarrow      \frac{M_j|\Psi\rangle}{\sqrt{\langle
  \Psi|M_j^\dag M_j|\Psi\rangle}},
\label{eq:collapse}
\end{equation}
with probability  $\langle\Psi|M_j^\dag M_j|\Psi\rangle$.
\end{description}
\bigskip
The last stage in the model is tantamount to controlling and directing
the   wavefunction  collapse   to  produce   an   outcome  probability
distribution that is closer (as quantified by trace distance, relative
entropy or any  other suitable distance measure) to  $P^U = \{p_j^U\}$
than $P=\{p_j\}$,  if FW  is available. If  FW is absent,  then Nature
alone   determines  the   outcome  probability   distribution,  which,
moreover, conforms to the Born rule.

When  $\sigma \ne  0$, the  resulting probabilities  of  outcomes will
violate the Born  rule.  In general, such violations  can give rise to
various   non-standard  effects,  among   them  violation   of  energy
conservation,   possibility  of   signaling   at  superluminal   speed
\cite{sri10}  and  the  possibility  of  counter-intuitive  models  of
computation   that   allow  efficient   solution   of  hard   problems
\cite{sri10}.  However, these non-standard effects will be confined to
a small sub-cellular region of the  brain, where it will not be easily
discernible from measurment  errors, decoherence effects, neural noise
and  statistical fluctuations.   Moreover, it  can be  facilely masked
behind the remaining features of  the brain's physiology, which can be
described  in  terms  of  deterministic, classical  mechanisms  (e.g.,
metabolic  changes leading  to  arousal potentials  in motor  neurons,
initiating familiar physical movements of the body).

Finally, it is worth  stressing that the extra-physical agency posited
to  determine  the outcome  of  wavefunction  collapse  should not  be
confused with hidden variables in  the sense of Bohm \cite{bohm}.  The
latter represents a device to turn QM into a deterministic theory, and
thus restore classical causality, whereas in our approach requires, FW
is a new form of causation.

\section{Neuroscientific implications, tests and applications \label{sec:neuro}}

In the conventional view, the  brain is a complex input-output device,
with apparently voluntary actions  arising from a mix of deterministic
and indeterministic mechanisms.  So it is an interesting question what
would constitute a falsifiable proof  of FW.  For example, in research
reported  in  Ref.  \cite{brembs},  tethered  fruit flies  (Drosophila
melanogaster) in a visually  impoverished environment had their flight
maneuvers recorded.  Lacking any external input, their random behavior
should have resembled a Poisson process.  However, the analysis showed
that  the temporal  structure of  fly behavior  follows  a L\'evy-like
probabilistic  behavior pattern.   At first  blush, it  seems possible
that this may be due to spontaneous behavior arising from (a primitive
version of) libertarian FW of  the fruit flies.  However, it turns out
that  the  observed  pattern  can  be simulated  via  intrinsic  noise
amplified by suitable nonlinearity \cite{brembs}.

Experiments that purport to  \textit{disprove} the existence of FW can
also  find alternate  explanations.  Based  on a  study  of volunteers
wearing scalp electrodes,  Libet and collaborators \cite{libet} showed
in 1983 that a `readiness potential' (RP) was detected a few tenths of
second before the subjects, in  their own reckoning, made the decision
to perform  an action (to flex  a finger or wrist).   Their result was
interpreted as indicating that the  motor cortex was preparing for the
action,  that unconscious  neural processes  predetermine  actions and
hence that  FW was illusory.  In  a recent comment  on the experiment,
Miller and Trevena \cite{trevena} asked  subjects to wait for an audio
tone before  making a decision to tap  a key or not.   If the activity
detected by Libet et al.  really  was the making of the decision prior
to any  conscious awareness of doing  so, then that  activity ought to
occur only  if the subjects decided  to act.  But  no such correlation
was found.  Miller and Trevena  conclude that the RP may only indicate
that  the brain  is  paying attention  and  does not  indicate that  a
decision has been made.

Under the  circumstances, one suggestion  would be to  trace backwards
along  a  deterministic observable  causal  chain  of  neurons to  the
specific  area  in the  motor  cortex  responsible,  perhaps a  single
neuron, that may  be considered as initiating a  free-willed action at
the  physiological level,  and  hence a  candidate  site for  carrying
signatures of the putative new  physics, such as the claimed deviation
from the Born rule.

Identifying  such single  neurons will  be difficult,  given  the high
density of neuronal packing,  and the weakness of excitatory synapses,
which would  make the role of  individual neurons in  the brain cortex
hard  to  identify.  Further,  the  extreme  complicatedness of  brain
dynamics  and  attendent decoherence,  measurement  errors, noise  and
statistical fluctuations,  will play an  adverse role.  A  study where
the twitching of a mouse's whisker has been traced to single pyramidal
neurons  in  the cortex  is  reported  in  Ref.  \cite{mouse}.   These
neurons  could  be  examined   for  unusual  behavior  that  might  be
considered  as appropriate  signatures, after  suitable  allowance has
been  made for noise  and innervations  from neighboring  neurons: for
example,  there  may be  fluctuations  in  the whisker-twitch  causing
potentials  that could  somehow  not be  ascribed  to any  neighboring
influences.  The   neuron  may  then  be  examined   carefully  for  a
sub-cellular  site  to which  the  initiation  of  the twitch  causing
potentials may be attributed.

Another  promising area  to  look  for evidence  suggestive  of FW  is
psychiatry  and neurology,  which are  startlingly revelatory  on how,
from   a    physiological   perspective,   the    human   mind   works
\cite{ramachandran}. The FUN\# model suggests that the decision making
ability  in  a human  is  affected,  possibly  leading to  anti-social
behavior, in  broadly three  \textit{different} ways: when  the limbic
system, or the frontal cortex, or  the pineal gland area (or any other
candidate UCM)  is impaired.  If in  some case such  behavior does not
coincide with impairment  to the limbic system or  the frontal cortex,
then  there  is  a  reasonable  case  to  attribute  the  behavior  to
diminished FW. Studying such cases,  if any, would help locate the UCM
and verify whether the pineal gland hypothesis holds good.

There are potential therapeutic  and medical applications based on the
above observations.  The FUN\#  model can help clarify what abnormalcy
of behavior  means and can  help classify abnormal depending  of which
faculty  is affected.   If  a particular  behavioral  disorder can  be
attributed  to  one  of  the  three above  causes,  then  the  therapy
prescribed to the  afflicted patient could also be  varied with better
efficacy. 

\section{Conclusions and Discussions \label{sec:end}}

A conservative,  Darwinian view  regarding the origin  of FW  would be
that it enables an organism to  deviate from the local gradient in the
brain's energy landscape,  and to better than locally  optimize in the
struggle  for self-presevervation,  which would  require improvisation
under novel situations;  once this was done, FW  brought along with it
potential by-products, like altruism, self-harm, etc., behaviors which
are  not necessarily  conducive  to self-preservation,  and hence  not
`intended' by the  evolutionary process that gave rise  to FW. Another
possibility is that FW is an  emergent phenomenon that is a product of
self-organization in complex quantum physical systems. These views are
not necessarily incompatible with the  FUN model. If anything, one can
be  adventurous  and  suggest  this extra-physical  agency  guides  or
conduces to  Darwinian evolution to make the  human physiology optimal
for the physical experience! For  example, it is known that the higher
the oxygen  concentration (up to  a point), the  better it is  for the
metabolism and life of organisms. It  is also known that the lower the
altitude, the  higher the  oxygen concentration.  If  there is  FW, an
organism in a mountainous area may feel the urge to explore farther by
climbing higher,  thereby improving its  chances of reaching  a valley
that is deeper, due to  which the oxygen concentration is higher, than
in the valley reached by following the local gradient.

Finally we offer our opinion on FW being, together with awareness,
aspects of consciousness.
To  be aware  of a  theory  is to  be able  to  talk about  it and  to
understand its implications and limitations. To this end, one requires
a meta-theory, which is equipped with a langauge in which to formalize
propositions  about the  theory.   An axiomatic  system  that is  rich
enough to encompass the meta-theory will, in general, be more powerful
than  the one  that axiomatizes  the theory.  An instance  of  this is
provided   by   G\"odel's  theorem   \cite{god},   which  provides   a
metamathematical proof of mathematically undecidable propositions.

If  FW did  not exist,  and the  behavior of  observers  were entirely
determined  by the  rules  of  the base  theory,  then the  observers'
algorithmic complexity  \cite{chaitin} would not be  greater than that
of the theory, making  them incapable of encompassing the meta-theory.
Equipped with  FW, human experimenters  can be {\it  meta-entities} in
the theory. It allows them to freely pose questions, perform tests and
draw inferences about the theory  as external agents whose choices are
not entirely determined by the  theory. From this perspective, FW
is an aspect of the consciousness of sentient observers that is
necessary to make them aware of a physical  theory such as
QM and to experience the (quantum) world as humans do.   



\begin{thebibliography}{99}
\bibitem{fw3} Wikipedia entry of free will.
\texttt{en.wikipedia.org/wiki/Free\_will}
\bibitem{suControl} A. Suarez. Quantum randomness can be controlled by
  free   will–-  a  consequence   of  the   before-before  experiment.
  arxiv:0804.0871.
\end{thebibliography}
\end{document}